\documentstyle[preprint,aps,epsf]{revtex}

\begin{document}
\draft \preprint{}
\title{Long range correlations in DNA : scaling properties and charge transfer efficiency}
\author{Stephan Roche${\ }^{1}$, Dominique Bicout${\ }^{2}$, 
Enrique Maci\'{a}${\ }^{3}$, Efim Kats${\ }^{2,4}$ }
\address{ ${\ }^{1 }$ Commissariat \`a l'\'Energie Atomique, DSM/DRFMC/SPSMS, 17 avenue des Martyrs, 38054 Grenoble, France\\
${\ }^{2}$Laue-Langevin Institute, 6 rue Jules Horowitz, BP 156, F-38042, Grenoble, France\\
${\ }^{3}$ Departamento de Fisica de Materiales, Facultad de Fisicas, Universidad Complutense, E-28040 Madrid, Spain\\
${\ }^{4}$L. D. Landau Institute for Theoretical Physics, RAS, Moscow, Russia}
\date{\today}
\maketitle
\begin{abstract} 
We address the relation between long range correlations and charge transfer efficiency in aperiodic artificial or genomic DNA sequences. Coherent charge 
transfer through the HOMO states of the guanine nucleotide is studied using 
the transmission approach, and focus is made on how the sequence-dependent 
backscattering profile can be inferred from correlations between base pairs.
\end{abstract}


During the past few years, the nature of {\it long range correlations} in DNA 
sequences has been the subject of intense debate\cite{DNA-corr1,DNA-corr2,DNA-corr3}. 
Scale invariant properties in complex genomic sequences with thousands of nucleotides 
have been investigated in particular with wavelet analysis \cite{DNA-corr2}, and 
have been argued to play crucial role in gene regulation and cell division. Besides, 
amongst the many physical, chemical or biological phenomena that might be inferred 
from sequence correlations, charge transfer properties deserve particular concern. 
Indeed, a precise understanding of DNA-mediated charge migration would have strong 
impact on the description of damage recognition process and protein binding, or 
in engineering biological processes \cite{DNAelec,Barton1}. 
The $\pi$-stacked array of DNA base pairs (bp) (made up from nucleotides: 
guanine $g$, adenine $a$, cytosine $c$, thymine $t$) provides an extended 
path to convey long range charge transport although dynamical motions of base pairs, 
or energetic sequence dependent heterogeneities, are expected to reduce   
long range efficiency. Photoexcitation experiments have unveiled that charge 
excitations can be transmitted between metallointercalators, preferentially 
through the guanine highest occupied molecular orbitals ($g$-HOMO) of 
the DNA bridge\cite{Barton1,DNAHT}. Such experiments and mesoscopic transport 
measurements on single artificial or genomic DNA sequences contacted in between 
metallic electrodes have also been the subject of intense and controversial debate 
\cite{DNA-TRS}. While accurate determination of absolute values 
of conductivity is important, characteristic sequence dependences of charge 
transport could provide valuable clues to mechanisms and biological functions of 
transport. Such issue has been up to now poorly addressed experimentally and theoretically. In that perspective, the possible role of long range correlations on 
electronic delocalization has been recently anticipated\cite{DNA-corrTRS}. In this Letter, 
the electronic transport properties are proven to be critically related to the nature and range of correlations.


Rescaling coefficients have been introduced as a useful measure of correlations 
in DNA sequences \cite{DNA-corr1}. It relies on the evaluation of the second 
moment of the fluctuations of sequence composition. 
The statistical method consists on constructing a mapping of the nucleotide 
sequence onto a walk. A DNA walk is initiated from the first to the last 
nucleotide of the sequence with the rule that the walker steps down 
$\left[v(i)=-1\right]$ if a purine ($a\,,g$) occurs at position 
$i$ along the sequence, whereas the walker steps up $\left[v(i)=+1\right]$ 
if a pyrimidine ($t\,,c$) occurs at position $i$. Given a nucleotide sequence 
of size $N$, the net displacement $x(n)$ of the nucleotide walker after $n$ 
steps is, $x(n)=\sum_{i=1}^{n}v(i)\:\:\:;\:\:\:1\leq n\leq N\:.$ Recently, 
Hurst's analysis \cite{HU50} was argued to be more reliable for determining 
the precise rescaling coefficients\cite{GRI98}. We thus follow the prescription of 
Hurst's analysis to construct adjusted variables as 
$X(m,k)=\Delta x(m,k)-\frac{k}{n}\,\Delta x(m,n)\:\:\:;\:\:\:1\leq k\leq n\:$ 
and define the range $S(m,n)$ for random walks of lengths $n$ as 
$S(m,n)=\max_{1\leq k\leq n}\left[X(m,k)\right]-
\min_{1\leq k\leq n}\left[X(m,k)\right]\:$. Now, the rescaled range 
function $R(n)$ is defined as \cite{HU50}, 
{\small
\begin{equation}
R(n)=\frac{\langle S(n)\rangle}{\sigma(n)}\propto n^H\:
\label{eq1}
\end{equation}
}
\noindent
where $\langle S(n)\rangle=\sum_{m=1}^{N-n}S(m,n)/(N-n)$ and $\sigma^2(n)$ 
is the standard deviation of $v(i)$ over walks of lengths 
$n$, and averaged over the entire sequence. The Hurst exponent $H$ of the 
process is then defined through the scaling in Eq.(\ref{eq1}). Interestingly, 
for short-ranged correlated random walk the exact result for the rescaled 
range function reads, 
$R(n)=\sqrt{[\pi n/2]}-1$ \cite{HU50,FE51}. Thus, $H=1/2$ for the ordinary Brownian 
motion. The existence of power-law behaviors suggests that there is 
no characteristic length scale associated with properties under consideration. It 
is clear at the first glance that DNA sequences are unlikely fully characterized 
by a single scaling exponent. One expects that the scaling behavior be different 
for different length scales of the sequence, i.e, the rescaling exponent is itself 
a function of the length scale $n$. In the case where a characteristic size 
$n_c$ can be defined, one may postulate that $R(n)$ is still described by the 
power-law in Eq.(\ref{eq1}), but with a scale dependent rescaling exponents $H(n)$ 
such that $H(n)=H_1$ for $1\leq n < n_c$ and $H(n)=H_2$ for $n\geq n_c$. 

In our study, we consider three sequences: a DNA sequence of the first completely 
sequenced human chromosome $22$ (Ch22) containing about $33.4\times10^{6}$ 
nucleotides 
entitled ${\rm NT}_{011520}$ retrieved from the National Center for Biothechnology 
Information (NCBI), a Random DNA sequence (where $a\,,c\,,t\,,g$ are evenly chosen 
probability $1/4$) and a Fibonacci Polygc quasiperiodic sequence constructed 
starting from a $g$-nucleotide as seed and following the inflation rule 
$g\rightarrow gc$ and $c\rightarrow g$. This gives successively  
$g\,,gc\,,gcg\,,gcggc\,,gcggcgcg\,,gcggcgcggcggc\,,\cdots$, for 
sequences of length $1$, $2$, $3$, $5$, $8$, $13$, $\cdots$, respectively, such that 
its characteristic self-similar order introduces correlations on broad scale range. 
The ratio $[\mbox{number of g}]/[\mbox{number of c}]$ approaches the golden mean 
value $(1+\sqrt{5})/2\simeq 1.618$ in the limit of an infinite sequence. The Random 
and Fibonacci sequences are used as prototypes of short-range (or uncorrelated) 
and strongly correlated systems, respectively. 



\begin{figure}
\epsfxsize=\linewidth
\epsfbox{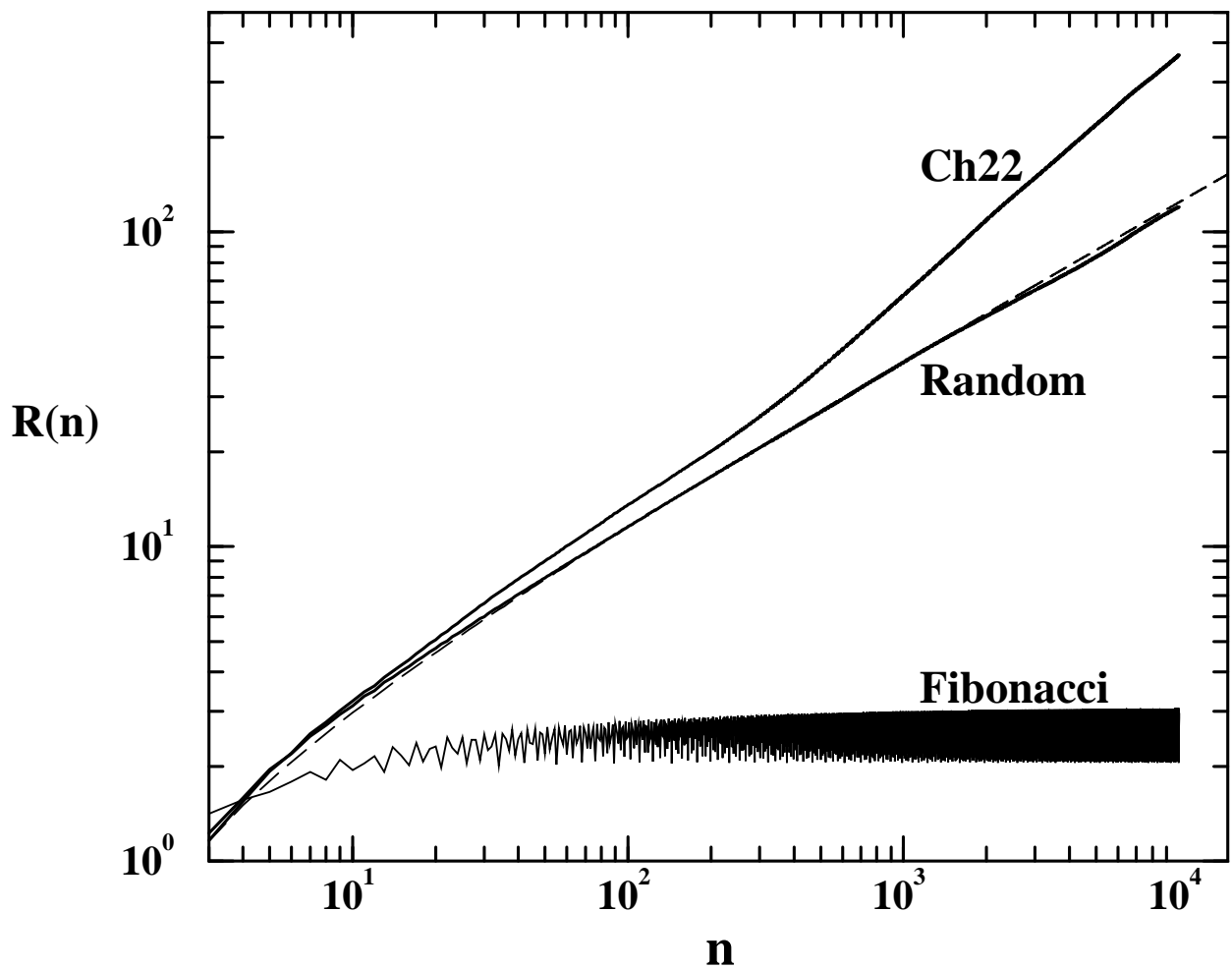}
\vspace{2.5pt}
\caption{Rescaled range function $R(n)$ versus $n$. Dashed line 
corresponds to $\sqrt{[\pi n/2]}-1$.}
\label{FIG1}
\end{figure}

The computed functions $R(n)$ for the three sequences described above are reported 
on Fig.~\ref{FIG1} and values of $H$ are summarized in Table~\ref{tb1}. It clearly 
appears from these calculations that the Random sequence is indeed uncorrelated 
following the $\sqrt{[\pi n/2]}$-law, whereas Fibonacci sequence is strongly correlated
with a ''ballistic behavior'' and correlations in Ch22 sequence 
exhibit a power-law behavior with a scaling exponent depending on the length scale.
The Ch22 sequence has long-range correlations characterized by Hurst exponents 
greater than $1/2$ (see Table~\ref{tb1}).
Given the huge amount of nucleotides of the Ch22 sequence, the physically relevant 
question seems rather to address to which extent charge transport can be efficient 
through the $g$-HOMO, in comparison with uncorrelated random or quasiperiodic 
sequences. To have some elements of response, we now turn to the examination of 
charge transfer properties in these sequences. To this end, we consider an 
effective tight-binding Hamiltonian describing the energetics of a hole 
located at nucleotide site $n$ \cite{Ratner,HDNA},
\begin{equation}
{\cal H} = \sum_{n}\varepsilon_{n} c_{n}^{\dagger}c_{n}
-\sum_{n}t_{0}(c_{n}^{\dagger}c_{n+1}+h.c.)\:
\label{eq2}
\end{equation}
where $c_{n}^{\dagger}$ ($c_{n}$) is the creation (annihilation) 
operator of a hole at site $n$. The hole site energies $\varepsilon_{n}$ are 
chosen according to 
the ionization potentials of respective bases \cite{HDNA}, $\varepsilon_{a}=8.24eV$, 
$\varepsilon_{t}=9.14eV$, $\varepsilon_{c}=8.87eV$, and $\varepsilon_{g}=7.75eV$, 
while the hopping integral, simulating the ${\rm \pi-\pi}$-stacking between adjacent 
nucleotides, is taken as $t_{0}=1eV$. The DNA sequences are further assumed to be 
connected to two semi-infinite electrodes whose energies  $\varepsilon_{m}$ are 
adjusted to simulate a resonance with the $g$-HOMO energy level, 
$\varepsilon_{m}=\varepsilon_{g}$, and with hopping integrals such that 
$t_{m}=t_{0}$. Note that ab-initio studies suggest that 
$t_{0}\sim 0.1-0.4eV$\cite{HDNA}, but the choice $t_{m}/t_{0}=1$ reduces 
backscattering of holes at the contact electrodes and allows for a larger 
accessible transmission spectrum and a better characterization of 
DNA's intrinsic conduction \cite{Ratner}. Sites comprised between 
$[-\infty,0]\cup [N+1,+\infty]$ belong to the leads, whereas sites $i=1,N$ are 
associated to the sequence of size $N$ under study. The transmission coefficients 
are computed using the transfer matrix formalism in which the time independent 
Schr\"{o}dinger equation is projected into a localized basis by properly accounting 
for the boundary conditions \cite{Landauer}. Let $\psi_{n}$ denotes the wavefunction 
with energy $E$ at site $n$, we obtain from Eq.(\ref{eq2}) the recurrent equation, 
{\small
$\left(\begin{array}{c}
\psi_{N+2} \\ 
\psi_{N+1}
\end{array}\right)=M_{N}
\left(\begin{array}{c}
\psi_{N+1} \\ 
\psi_{N}
\end{array}\right)=M_{N}\cdots M_{1}
\left(\begin{array}{c}
\psi_{1} \\ \psi_{0} 
\end{array}\right)\:,
$}
where $M_n$ is a $2\times 2$ matrix with elements $M_{n}(1,1)=(E-\varepsilon_{n})/t_{n+1}$, 
$M_{n}(1,2)= -t_{n}/t_{n+1}$, $M_{n}(2,1)=1$ and $M_{n}(2,2)=0$. The transmission coefficient 
$T_{N}(E)$, that gives the fraction of tunneling electrons transmitted through the N-site DNA, 
is related to the Landauer resistance as $(h/2e^{2})[1-T_{N}(E)]/T_{N}(E)$, where 
$h/2e^{2}$ is the quantum resistance and \cite{Landauer},
{\small 
\begin{eqnarray}
T_{N}(E)=\left[4-\frac{(E-\varepsilon_{m})^{2}}{t^{2}_{m}}\right]\left/
\left\{-\frac{(E-\varepsilon_{m})^{2}}{t^{2}_{m}}\,({\cal P}_{12}{\cal P}_{21}+1) 
\right.\right.  & &  \nonumber \\
\left.+\frac{(E-\varepsilon_{m})}{t_{m}}\,({\cal P}_{11}-{\cal P}_{22})({\cal P}_{12}-{\cal P}_{21})
+{\sum_{i,j=1,2}} {\cal P}_{ij}^{2} + 2 \right\}\: & &
\label{eq4}
\end{eqnarray}}
\noindent 
with ${\cal P}=M_{N}M_{N-1}....M_{1}$. For a given energy, $T_{N}(E)$ reflects the level 
of backscattering events in the hole transport through the sequence. As metallic leads are 
adjusted to the $g$-HOMO energy level, the hole transport will experience a sequence 
dependent contribution of backscattering according to the distribution of $c$, $t$, 
and $a$ potential barriers over the length scale of the sequence. To compare transmission 
properties of different chains, the behavior of the Lyapunov coefficient, 
$\gamma_{N}(E) = \frac{1}{2N} \ln(T_{N}(E))$, is also calculated. $\gamma_{N}(E)$ 
has been extensively investigated to sort out the main features of complex localization 
patterns \cite{Ly,Ly-qp}. For systems with uncorrelated disorder, $\gamma_{N}(E)$ 
provides the localization length $\xi(E)=1/[\lim_{N\to\infty}\gamma_{N}(E)]$. 
In presence of scale invariance properties, the underlying structure of 
$\gamma_{N}(E)$ reflects the self-similarity of the spectrum \cite{Ly-qp}. 

\begin{figure}
\center
\epsfxsize=\linewidth
\epsfbox{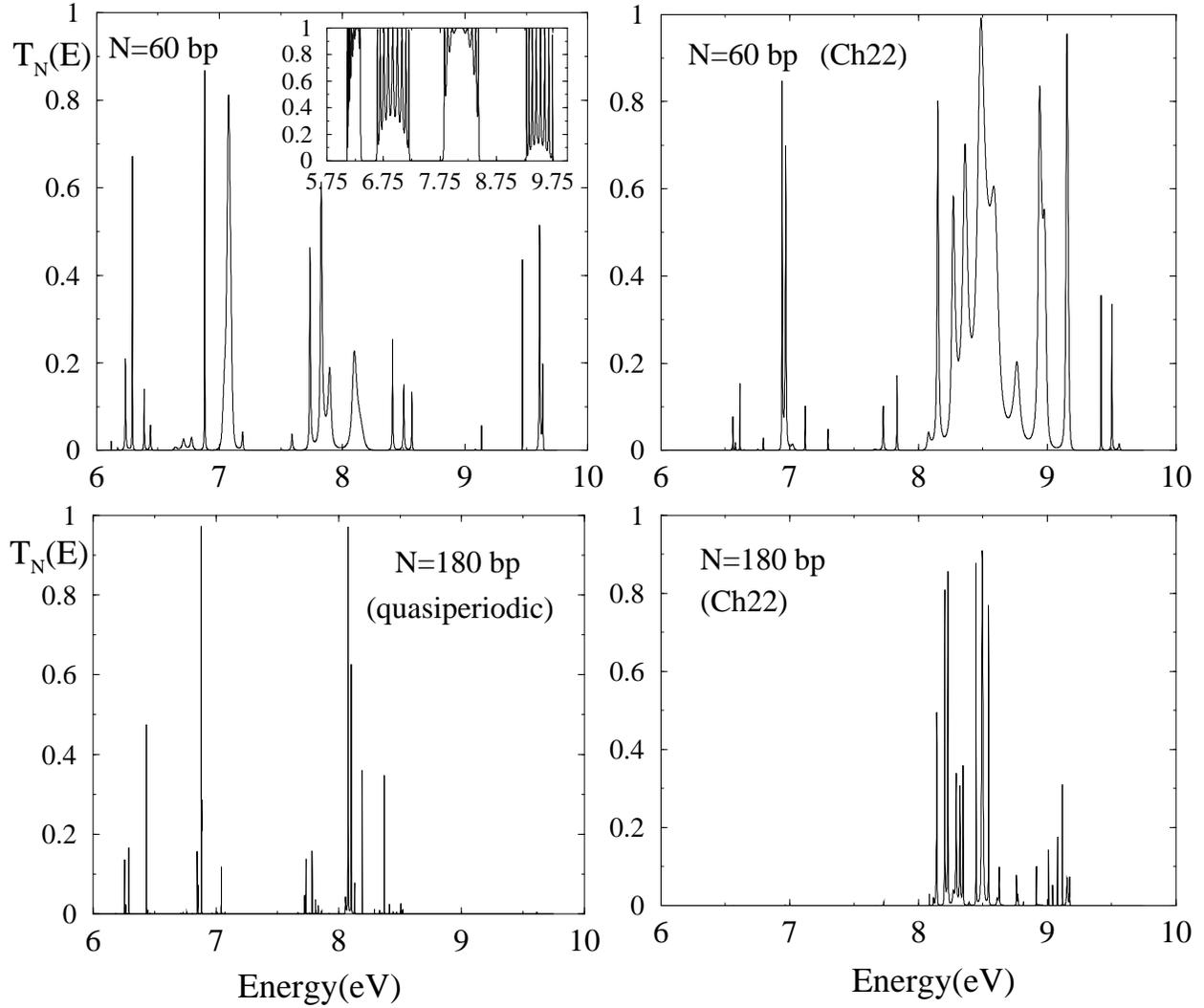}
\vspace{5pt}
\caption{Transmission coefficient for Fibonacci Polygc quasiperiodic (left frames) 
and Ch22-based sequences (right frames). Inset: $T_N(E)$ in Eq.(\ref{eq5}) for 
a periodic approximant of length $N=50$ bp.}
\label{FIG2}
\end{figure}
\begin{figure}  
\center
\epsfxsize=\linewidth
\epsfbox{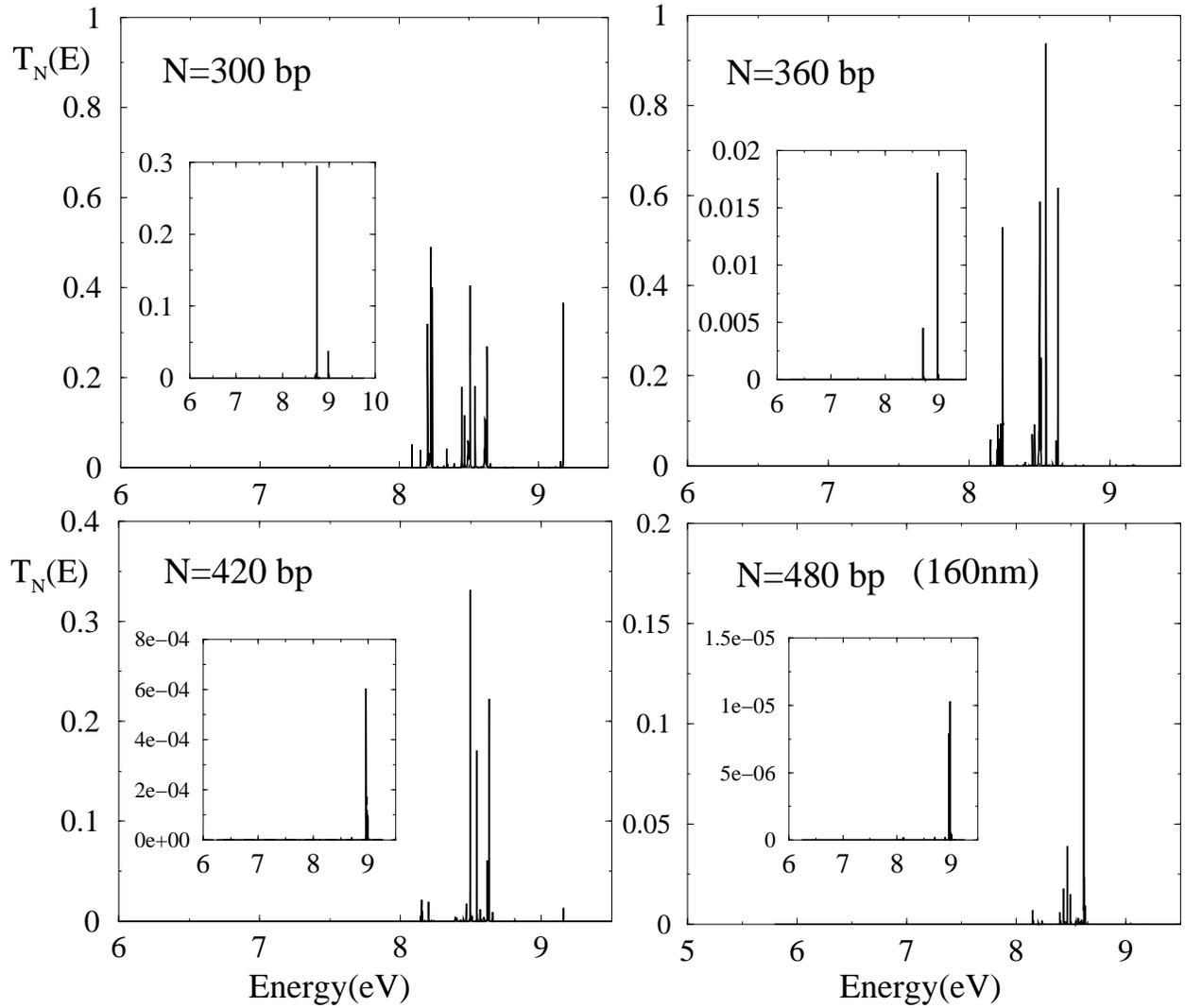}
\vspace{5pt}
\caption{Transmission coefficient for Ch22-based sequences (main frames) 
and typical results (over about 50 sequences) for uncorrelated DNA random chains (insets) with same number of nucleotides.}
\label{FIG3}
\end{figure}
\begin{figure} 
\epsfxsize=\linewidth
\epsfbox{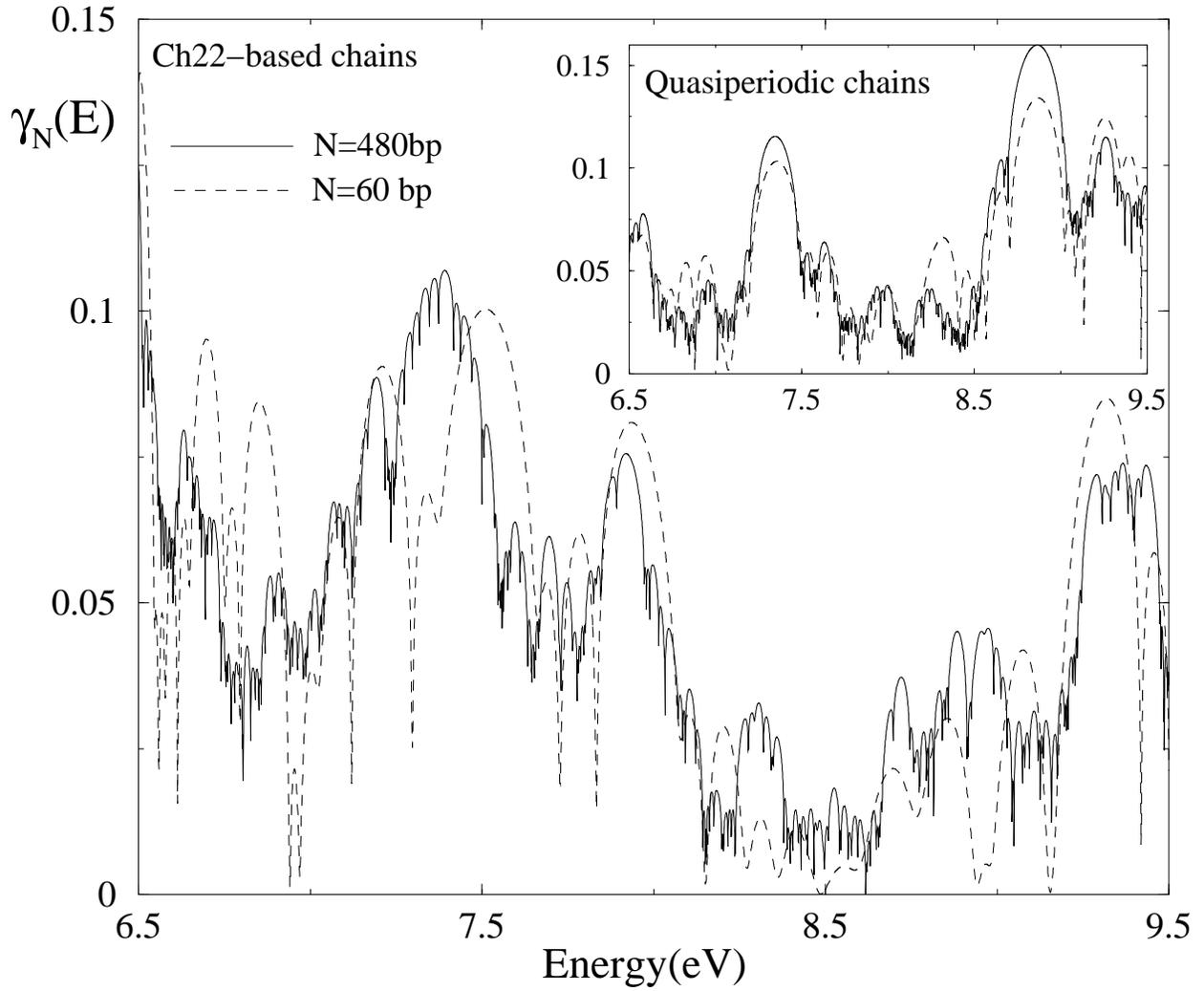}
\vspace{5pt}
\caption{Lyapunov coefficient for Ch22-based (main frame) and 
Fibonacci Polygc quasiperiodic sequences (inset).}
\label{FIG4}   
\end{figure}

Following our analysis on correlations, the $T_{N}(E)$ for the three sequences of 
Table~\ref{tb1} have been computed, varying the sequence length. The random and Fibonacci quasiperiodic 
based sequences are generated starting from the first nucleotide of the sequence up to 
$N$ bp, while the Ch22-based sequences are constructed by starting from the 
bp=15000 of the full Ch22 sequence and then extracting the first $N$ bp, namely 
$agggcatcgctaacgaggtcgccgtccaca$ $gcatcgctatcgaggacaccacaccgtcca$ for $N=60$ bp.  
Figures \ref{FIG2} and \ref{FIG3} present the comparison of $T_{N}(E)$ between the 
quasiperiodic and Ch22 sequences and between uncorrelated random DNA and Ch22 
sequences, respectively, with the same number of bp. Lyapunov coefficients for 
quasiperiodic and Ch22-based sequences are also displayed in Fig.~\ref{FIG4}. 

General trends of Figs.~\ref{FIG2} and \ref{FIG3} are that $T_{N}(E)$ is characterized 
by an energy spectrum of resonant peaks with high transmission. As the sequence length 
increases, much less states will present good transmittivity, due to the progressive 
fragmentation of the spectrum, although several peaks with high transmission remain at 
certain energy values, and new ones may appear. For Fibonacci and Ch22-based sequences, these resonant energies are robust enough to persist against backscattering effects 
due to interspersed bases along the sequence.  This point is illustrated in Fig.2 and Fig.3 where one observes that Fibonacci (resp. Ch22-based sequences) of $180$ bp (resp.360 bp) exhibit 
states with better transmission properties than those present in a $60$ bp (resp. 300bp) long sequence. In addition, $\gamma_{N}(E)$ shown in Fig.~\ref{FIG4} illustrates intrinsic properties of the two correlated sequences albeit of different nature. Indeed, the 
series of main elliptic bumps found in the Fibonacci sequence with $60$ bp are 
reproduced in the $480$ bp sequence, which present additional features associated 
with the partitioning of spectrum. While self-similarity fully characterizes the 
quasiperiodic sequence, the scaling properties in Ch22 rely on totally different 
kind of long range correlations, with no hints of self-similar patterns. 


In contrast, the fragmentation of the spectrum strongly affects the transmittivity 
of the uncorrelated random sequence. All resonant states (when any) are evenly 
affected and the corresponding transmission decreases as the sequence length gets 
longer. From a statistical analysis over many random sequences, it clearly appears 
that Ch22-based sequences exhibit much higher charge transfer efficiency over 
much longer distances in comparison with uncorrelated random sequences.


Nevertheless, to improve our understanding and gain some physical insights about 
characteristic features exhibited by these sequences, we now focus on quasiperiodic 
sequences since it has been shown that the global structure of the electronic spectrum 
of such chains can be obtained in practice by considering very short periodic 
approximants to infinite quasiperiodic chains\cite{Ly-qp}. These sequences are 
characterized by long range correlations that manifest themselves on electronic 
properties in terms of power-law localization of eigenstates in the thermodynamic 
limit or power-law increase of Landauer resistance in finite samples\cite{Ly-qp}. 
For this purpose, we consider a periodic approximant whose unit cell is  $gcggc$. 
The corresponding dispersion relation of this approximant is given by, 
{\small $2t_0^{5}\cos (5q)=\allowbreak \left( E-\varepsilon _{g}\right) ^{3}
\left(E-\varepsilon _{c}\right) ^{2}-t_0^{2}\left( E-\varepsilon _{g}\right) 
\left(E-\varepsilon _{c}\right) \left( 5E-4\varepsilon _{g}-\varepsilon_{c}\right) 
+t_0^{4}\left(5E-3\varepsilon _{g}-2\varepsilon _{c}\right)$}. The energy 
spectrum of the $gcggc$ approximant is composed of three broad bands (of 
bandwidth $\simeq 0.5-0.6$ eV) centered at the energies 
$E_{2}=6.\,\allowbreak 915$ eV, $\allowbreak E_{3}=8.143$ eV and 
$E_{4}=9.\,\allowbreak 527$ eV, plus two narrower bands (of bandwidth 
$\simeq 0.25$ eV) located at the edges of the spectrum at $E_{1}=6.191$ eV 
and $E_{5}=10. 213$\ eV. These analytical results allow us to properly assign the 
different resonant peaks appearing in the spectrum of the transmission coefficient 
(shown in the inset in Fig.~\ref{FIG2}) in respect to the four main sub-bands of 
the spectral window $[5.75$, $9.75$ eV]. States belonging to the broader central 
bands around $E_{2}=6.915$ eV and $E_{3}=8.143$ eV turn out to be very robust to 
the progressive fragmentation of the energy spectrum. Accordingly, one is 
tempted to conclude from the simple inspection of Fig.~\ref{FIG2} (left frames) 
that these states should exhibit good transport properties even in the 
thermodynamic limit. To further substantiate such an assertion, we consider in 
addition the transmission coefficient corresponding to the $gcggc$ approximant, 
{\small 
\begin{equation}
T_{N}(E)=\left[ 1+q(x,y)U_{\frac{N}{5}-1}^{2}(w)\right] ^{-1}\:
\label{eq5}
\end{equation}
} 
\noindent where $x=(E-\varepsilon _{c})/2t_0$, $y=(E-\varepsilon_{g})/2t_0$, 
$w=16x^{2}y^{3}-16xy^{2}-4yx^{2}+3y+2x$ the $U_{n-1}(w)$ is a
Chebyshev polynomial of the second kind, and $q(x,y)\equiv
A^{2}/(1-y^{2})+B^{2}-1$ with $A\equiv
-24xy^{3}-16x^{2}y^{2}+6xy+2x^{2}+32x^{2}y^{4}+4y^{4}+y^{2}$ and 
$B\equiv 32x^{2}y^{3}-8x^{2}y-24xy^{2}+4y^{3}+3y+2x.$ The resonance 
condition then reads, $q(x,y)U_{\frac{N}{5}-1}^{2}(w)=0$, while the condition
$q(x,y)\equiv 0$ yields $E_{l}=4.\,\allowbreak 317$ eV (which does not
belong to the spectrum) and $E_{u}=10.\,\allowbreak 158$ eV (located near
the center of the uppermost band, which is not included in our spectral
window). On the other hand, the roots of the Chebyshev polynomial label a
full transmission peak series according to the relationship $w=\cos (5k\pi/N)$ 
with $k=0,...,N$. This is illustrated in the inset of Fig.~\ref{FIG2} 
(top-left) where one observes oscillations in the energy dependence of 
the transmission curve for a sequence $cgccg$ with 10 units. By a deeper 
analysis, we find that Fibonacci quasiperiodic sequences as long as $160$ nm 
i.e., $\sim 450$ bp will still allow for nearly resonant transmission around two 
specific energies $E_{2}\simeq 6.9$ eV and $E_{3}\simeq 8.1$ eV. 



In summary when compared with uncorrelated sequences, long range correlations in aperiodic DNA sequences seem to induce coherent charge transfer over longer length scales. Such feature has been illustrated in particular in Chromosome 22-based sequences. Given that the nature of long range correlations differs in coding versus non-coding regions of genomic DNA\cite{DNA-corr3}, one should further elaborate on a more systematic study of charge transport in genomic DNA.

\vfill\eject

\begin{table} 
\begin{tabular}{lcccc} 
Sequence & $N$ & Purines & \multicolumn{2}{c}{$H(n_c=300)$} \\ \cline{4-5}
 &  &   & $H_1$ & $H_2$ \\ \hline
Ch22 & $182617$ & $91029$  & $0.60$  & $0.75$ \\
Random  & $182617$ & $91118$ &  $0.50$ & $0.50$ \\
Fibonacci  & $46368$ & $28657$ &  $0.085$ & $0.011$ \\ 
\end{tabular} 
\caption{Hurst exponents calculated from data in Fig. 1.}
\label{tb1} 
\end{table} 

\vfill\eject

\end{document}